# SFTP: A Secure and Fault-Tolerant Paradigm against Blackhole Attack in MANET

Jitendra Kumar Rout
Department of CSE
NIT, Rourkela

Sourav Kumar Bhoi
Department of CSE
NIT, Rourkela

Sanjaya Kumar Panda
Department of CSE
NIT, Rourkela

## ABSTRACT
Security issues in MANET are a challenging task nowadays. MANETs are vulnerable to passive attacks and active attacks because of a limited number of resources and lack of centralized authority. Blackhole attack is an attack in network layer which degrade the network performance by dropping the packets. In this paper, we have proposed a Secure Fault-Tolerant Paradigm (SFTP) which checks the Blackhole attack in the network. The three phases used in SFTP algorithm are designing of coverage area to find the area of coverage, Network Connection algorithm to design a fault-tolerant model and Route Discovery algorithm to discover the route and data delivery from source to destination. SFTP gives better network performance by making the network fault free.

## 1. INTRODUCTION
Mobile Ad-Hoc Network is a self-organizable and self-configurable network. This allows the devices and the people to interconnect with no hindrance in the areas of no communication infrastructure [9]. MANET has many unique properties like movement of host frequently, no cellular infrastructure, multihop communication links, data transmission through the intermediate nodes etc. MANET is nowadays used in many applications like personal area networking, military applications, civilian environments, emergency operations etc [2], [7]. The architecture of MANET consists of applications and middleware layer, networking layer, enabling technology layer and consist of cross layer issues. The major issues which affect the performance of the MANET are Medium Access Scheme, routing, multicasting, service discovery, transport layer protocol, pricing scheme, QoS provisioning, self organized, security, energy management, addressing, service discovery, scalability, deployment etc.

These are the main security issues in MANET which is an important for better network performance. The attacks are generally classified as active attack and passive attack. Active attack is the more dangerous attack in the network which modifies the data and passive attack is the attack without any modification of data. In MANET routing protocol plays the most important role in route discovery and data transmission. There are three types of routing protocols. First is the Proactive routing protocol which maintains the information continuously by a table containing information about the source and the destination and this table is regularly updated. DSDV, FSR, OLSR, STAR etc. are some of the proactive routing protocols. Reactive routing protocol is the on demand driven protocol which is used when requested and then only it will start the route discovery. AODV, DYMO, TORA etc. are some of the reactive routing protocols. Hybrid routing protocol is the one which is the combination of both proactive and reactive. Here, the network is divided into zones and in each zone it uses a different protocol. ZRP and ZHLS are the protocols used in hybrid routing protocols. So, these are some of the outline about MANET. In this paper, we have proposed a Secure Fault-Tolerant Paradigm (SFTP) which checks the Blackhole attack in the network [3], [13], [14]. The three phases used in SFTP algorithm are designing of coverage area to find the area of coverage, Network Connection algorithm to design a fault-tolerant model and Route Discovery algorithm to discover the route and data delivery from source to destination. SFTP gives better network performance by making the network fault free from different types of attack [10].

The organization of the paper is as follows: section II presents the preliminaries, section III discusses about the previous work done, section IV presents the proposed SFTP algorithm. Section V presents the modeling and observation of SFTP model and section VI presents the conclusion.

## 2. PRELIMINARIES
### 2.1 Routing Protocols
Routing protocols are mainly used to deliver the data securely and efficiently and for route discovery and discovers the network topology [1]. It dynamically maintains the route between the nodes. These protocols are essential because of the mobility of the nodes. There are three types of routing protocols. First is the Proactive routing protocol which maintains the information continuously by a table containing information about the source and the destination and this table is regularly updated. DSDV, FSR, OLSR, STAR etc. are some of the proactive routing protocols. Reactive routing protocol is the on demand driven protocol which is used when requested and then only it will start the route discovery. AODV, DYMO, TORA etc. are some of the reactive routing protocols. Hybrid routing protocol is the one which is the combination of both proactive and reactive. Here, the network is divided into zones and in each zone it uses a different protocol. ZRP and ZHLS are the protocols used in hybrid routing protocols. Here, we have used the concept of DSR routing protocol.

### 2.2 Attacks in MANET
Security issues in MANET are a main issue of study nowadays [7]. To achieve confidentiality, authenticity, integrity and non-repudiation we have to check the different type of attacks on the network. There are two types of attacks on are passive and the other is an active attack [8]. Passive attack consists of traffic analysis and monitoring. Active attacks are in different layers like MAC layer attacks, Network Layer attack, Transport layer attack, Application layer attack and other attacks are also there. Network layer attacks are like Wormhole, Blackhole, Sybil, Byzantine, fabrication etc. MAC Layer attacks are like jamming. The transport layer attack is like session hijacking, SYN flooding etc. Application layer attack is like data corruption and other attacks consists of Jellyfish, Spoofing, flooding, Denial of service, Grayhole etc. In this section we have used the Blackhole attack which means dropping the packet in the network by creating network disruption, congestion, delay etc [3], [13], [14].

## 2.3 Assumptions
SFTP algorithm is based on many assumptions. We have considered a MANET where each node is responsible for communication. The network is a location aware network. The node having high communication links is the busy node and the active node in the network. So, we have assumed the node as the high priority node and the node having low communication links is the low priority node. If a node fails then we connect the adjacent nodes of the failed node to tolerate the faults in the network.

## 2.4 Performance Metrics
The performance metrics which show the network performance are End-to-End Delay, Packet delivery rate and Throughput. These are the parameters which show the efficiency of the network.

## 3. RELATED WORK
Many researchers have been going under security issues in MANET. Tamilselval et al. Proposed a method to prevent a Blackhole attack in MANET using AODV protocol [15]. Balakrishnan et al. explained about the identification of malicious nodes by fellowship model [4]. Burmester et al. found a secure route discovery algorithm endairA [6]. Bhalagi et al. proposed an approach in which they classify the nodes in 3 categories based on their behavior to combat Blackhole attack [5]. Zhang et al. proposed a detection method to check Blackhole attack based on checking the sequence number in the Route Reply message [17] by generating a new message. Medadian et al. proposed a method to combat against Blackhole attack by using negotiation with the neighbors claiming to have a route to the destination [11]. Papadimitratos et al. found a method for secure data communication by using SMT and SSP [12].

## 4. PROPOSED SFTP ALGORITHM
### 4.1 Description
In our proposed SFTP algorithm we have proposed a model for tolerating and resisting the attacks and faults by constructing an efficient secure MANET. In our algorithm there are three phases in which first we find the coverage area of the network. Secondly, we connect the nodes by the Node Connection algorithm in case of node failure (inactive node). This is done by sending ping messages to all the nodes in the network to check whether the nodes are working. This is done before data communication between Source (S) and Destination (D). If we are unable to get the response then we are sure that the node is totally failed. Then we connect the adjacent nodes of the failed node to tolerate the faults in the network by using a Node Connection algorithm. Then we send the data from Source to Destination. Thirdly, we use the Route Discovery algorithm to decide the minimum cost route to reach the destination by resisting the malicious node attacks in the network. This is done by using DSR algorithm.

### 4.2 Description
Our proposed SFTP algorithm consists of three phases. The three phases are:

#### 4.2.1 Designing of Coverage Area
This is the first phase of SFTP algorithm. This is done to transform the network to a Wi-Fi range (secure range). This algorithm is mainly done for finding the coverage area.

---------------------------------------------------------------------------
**Algorithm 1: Designing of Coverage Area**
---------------------------------------------------------------------------
1. Construct the adjacency matrix A[i][j] by entering the range (weight).
2. By using the threshold value the Wi-Fi range is created.
3. Update the adjacency matrix as $A_U[i][j]$ after applying the threshold.

#### 4.2.2 Node Connection Algorithm
After finding the coverage area of the network we use the Node Connection Algorithm. This algorithm is to check the failure of the node before transmission of data from source to destination. This is done by sending ping messages to the nodes. If we get the responses then we come to know that the node is active otherwise it is inactive and failed. This is done before the data communication occurs between the source and the destination. Then we connect the nodes for tolerating the faults in the network by connecting the adjacent nodes of the failed node. This is done by connecting the nodes of high priority to low priority. The node having a high communication path has highest priority. If the degree of the nodes is same then we connect the nodes according to the convention. After this we get a total fault tolerant network.

---------------------------------------------------------------------------
**Algorithm 2: Node Connection Algorithm**
---------------------------------------------------------------------------
1. Send Ping messages to all the nodes in the network.
2. if (response comes from the nodes)
       The node is active.
    else
       The node is inactive.
3. Start the communication using the active nodes in the network.
4. Create Node Connection Table by prioritizing the nodes.
5. Connect the adjacent nodes of the failed node.
6. if the nodes have same degree then connect according to convention
7. Fault-tolerant network is created.

#### 4.2.3 Route Discovery Algorithm
After creating the fault-tolerant network we start the communication from source to destination. We know that, after applying the Node connection algorithm we get active nodes. In the active nodes many malicious nodes are also present. These adversary nodes may drop the packets ( Blackhole Attack) in the network by creating delay, packet loss, congestion, non optimal paths etc. So, to check the Blackhole Attack, we have proposed a simple Route Discovery Algorithm using Dynamic Source Routing (DSR) because this is the simple routing protocol with low route discovery overhead [16].

---
**Algorithm 3: Route Discovery Algorithm**
---
1. Start route discovery
2. Source node (S) floods the RREQ packet in the network in search for the destination node (D)
3. Broadcasting occurs
4. Append the list of identifiers of the RREQ
5. Each node getting RREQ floods the packet again to its neighboring nodes
6. The node which floods the packet will not further flood the packet
7. The transmission of two nodes may collide
8. After reaching the D node, no further RREQ packet transfer occurs
9. Record the list of identifiers of the RREQ
10. After getting the first RREQ, the D node sends RREP to the S node through that route
11. Record all the RREQ requests and the routes
12. The RREP is sent by reversing the route appended
13. RREP includes the route from source to destination which was received by the node
14. The route is discovered
15. Data delivery starts from source to destination
16. if (the packet send is not delivered to the other node in that route)
     {
        Packet Dropping Attack (Blackhole Attack) occurs in the node and it is a malicious node and packet transmission will not occur through that node
     }
     else if (the packet is sent with some delay)
     {
        Transmission occurs through that node and suddenly the timestamp of the packet is matched to know whether that node is a malicious node or not
     }
     else
     {
        Data delivery continues
     }

## 5. MODELLING & OBSERVATION
Here we have described the model by an illustration. In this example we have considered a MANET. Then, we find the coverage area by thresholding the range values. Then we apply the Node Connection Algorithm to get the fault-tolerant network. After this we apply the route discovery algorithm to send the packet from source to destination securely by resisting the Blackhole Attack.

### 5.1 Illustration
Firstly, we have considered a Mobile Ad Hoc Network. The steps for SFTP algorithm is as follows:

#### 5.1.1 Step 1
Find the network coverage area of the MANET shown in figure 1.

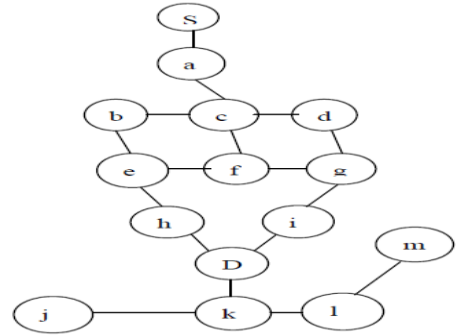

**Fig 1: Mobile Ad Hoc Network**

#### 5.1.2 Step 2
**Table 1. Adjacency Matrix**

|   | a | b | c | d | e | f | g | h | i | j | k | l | m | S | D |
|---|---|---|---|---|---|---|---|---|---|---|---|---|---|---|---|
| a | 0 | 0 | 1 | 0 | 0 | 0 | 0 | 0 | 0 | 0 | 0 | 0 | 0 | 1 | 0 |
| b | 0 | 0 | 2 | 0 | 2 | 0 | 0 | 0 | 0 | 0 | 0 | 0 | 0 | 0 | 0 |
| c | 1 | 2 | 0 | 2 | 0 | 2 | 0 | 0 | 0 | 0 | 0 | 0 | 0 | 0 | 0 |
| d | 0 | 0 | 2 | 0 | 0 | 0 | 2 | 0 | 0 | 0 | 0 | 0 | 0 | 0 | 0 |
| e | 0 | 2 | 0 | 0 | 0 | 2 | 0 | 3 | 0 | 0 | 0 | 0 | 0 | 0 | 0 |
| f | 0 | 0 | 2 | 0 | 2 | 0 | 2 | 0 | 0 | 0 | 0 | 0 | 0 | 0 | 0 |
| g | 0 | 0 | 0 | 2 | 0 | 2 | 0 | 0 | 1 | 0 | 0 | 0 | 0 | 0 | 0 |
| h | 0 | 0 | 0 | 0 | 3 | 0 | 0 | 0 | 0 | 0 | 0 | 0 | 0 | 0 | 3 |
| i | 0 | 0 | 0 | 0 | 0 | 0 | 1 | 0 | 0 | 0 | 0 | 0 | 0 | 0 | 1 |
| j | 0 | 0 | 0 | 0 | 0 | 0 | 0 | 0 | 0 | 0 | 4 | 0 | 0 | 0 | 0 |
| k | 0 | 0 | 0 | 0 | 0 | 0 | 0 | 0 | 0 | 4 | 0 | 1 | 0 | 0 | 4 |
| l | 0 | 0 | 0 | 0 | 0 | 0 | 0 | 0 | 0 | 0 | 1 | 0 | 3 | 0 | 0 |
| m | 0 | 0 | 0 | 0 | 0 | 0 | 0 | 0 | 0 | 0 | 0 | 3 | 0 | 0 | 0 |
| S | 1 | 0 | 0 | 0 | 0 | 0 | 0 | 0 | 0 | 0 | 0 | 0 | 0 | 0 | 0 |
| D | 0 | 0 | 0 | 0 | 0 | 0 | 0 | 3 | 1 | 0 | 4 | 0 | 0 | 0 | 0 |

Find the adjacency matrix table and updated adjacency matrix table by applying the threshold value as 4 shown in table 1 and table 2 respectively. D connected to k get splited. We have taken the nodes up to D (j, k, l, m are not taken), shown in figure 2.

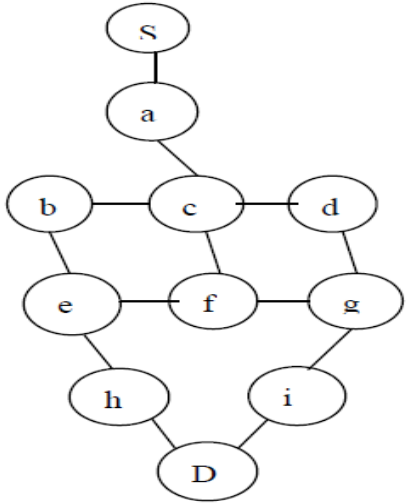

**Fig 2: MANET after thresholding with value 4**

### 5.1.3 Step 3
**Table 2. Updated Adjacency Matrix**

|   | a | b | c | d | e | f | g | h | i | j | k | l | m | S | D |
|---|---|---|---|---|---|---|---|---|---|---|---|---|---|---|---|
| **a** | 0 | 0 | 1 | 0 | 0 | 0 | 0 | 0 | 0 | 0 | 0 | 0 | 0 | 1 | 0 |
| **b** | 0 | 0 | 2 | 0 | 2 | 0 | 0 | 0 | 0 | 0 | 0 | 0 | 0 | 0 | 0 |
| **c** | 1 | 2 | 0 | 2 | 0 | 2 | 0 | 0 | 0 | 0 | 0 | 0 | 0 | 0 | 0 |
| **d** | 0 | 0 | 2 | 0 | 0 | 0 | 2 | 0 | 0 | 0 | 0 | 0 | 0 | 0 | 0 |
| **e** | 0 | 2 | 0 | 0 | 0 | 2 | 0 | 3 | 0 | 0 | 0 | 0 | 0 | 0 | 0 |
| **f** | 0 | 0 | 2 | 0 | 2 | 0 | 2 | 0 | 0 | 0 | 0 | 0 | 0 | 0 | 0 |
| **g** | 0 | 0 | 0 | 2 | 0 | 2 | 0 | 0 | 1 | 0 | 0 | 0 | 0 | 0 | 0 |
| **h** | 0 | 0 | 0 | 0 | 3 | 0 | 0 | 0 | 0 | 0 | 0 | 0 | 0 | 0 | 3 |
| **i** | 0 | 0 | 0 | 0 | 0 | 0 | 1 | 0 | 0 | 0 | 0 | 0 | 0 | 0 | 1 |
| **j** | 0 | 0 | 0 | 0 | 0 | 0 | 0 | 0 | 0 | 0 | 0 | 0 | 0 | 0 | 0 |
| **k** | 0 | 0 | 0 | 0 | 0 | 0 | 0 | 0 | 0 | 0 | 0 | 1 | 0 | 0 | 0 |
| **l** | 0 | 0 | 0 | 0 | 0 | 0 | 0 | 0 | 0 | 0 | 1 | 0 | 3 | 0 | 0 |
| **m** | 0 | 0 | 0 | 0 | 0 | 0 | 0 | 0 | 0 | 0 | 0 | 3 | 0 | 0 | 0 |
| **S** | 1 | 0 | 0 | 0 | 0 | 0 | 0 | 0 | 0 | 0 | 0 | 0 | 0 | 0 | 0 |
| **D** | 0 | 0 | 0 | 0 | 0 | 0 | 0 | 3 | 1 | 0 | 0 | 0 | 0 | 0 | 0 |

Creating the Node Connection table is shown in Table 3 in case of node failure to achieve fault tolerance.

**Table 3. Node Connection Table**

| Node | Communication Links | Priority |
|---|---|---|
| a | 2 | 4 |
| b | 2 | 5 |
| c | 4 | 1 |
| d | 2 | 6 |
| e | 3 | 2 |
| f | 3 | 3 |
| g | 3 | 4 |
| h | 2 | 7 |
| i | 2 | 8 |
| S | 1 | 10 |
| D | 2 | 9 |

### 5.1.4 Step 4
Apply the route discovery algorithm to the fault-tolerant network. Figure 3 shows the flooding of RREQ by S. Figure 4 shows the flooding of RREQ by a. Figure 5 shows the flooding of RREQ by c. Figure 6 shows the flooding of RREQ by b, d, f. Node d is the malicious node with Blackhole attack and supports in the route discovery to give an optimal path (advertises to give optimal path). The process continues and first RREQ reached D, is shown in figure 7.

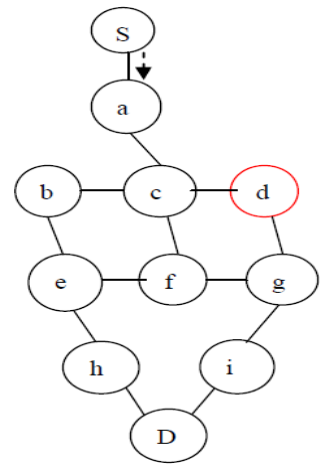

**Fig 3: Flooding of RREQ by S**

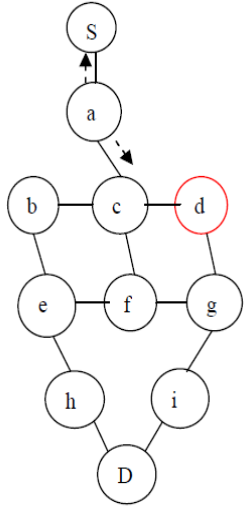

**Fig 4: Flooding of RREQ by a**

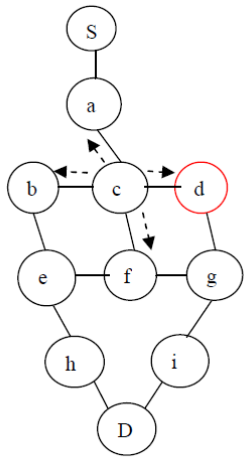

**Fig 5: Flooding of RREQ by c**

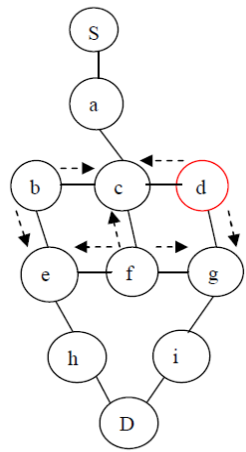

**Fig 6: Flooding of RREQ by b, d, f**

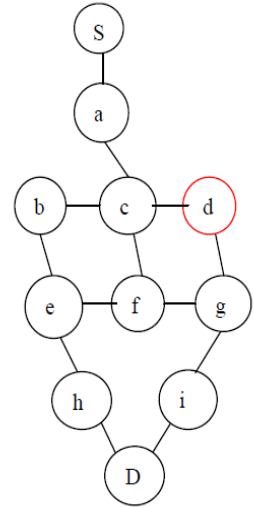

**Fig 7: RREQ requests are completed**

### 5.1.5 Step 5
Recording the list of identifiers of the first RREQ when received at D and then D sends the RREP to the S node by reversing the route appended. The list of identifiers is found to be [S, a, c, d, g, i, D]. Figure 8 shows RREP in a reverse manner to S from D.

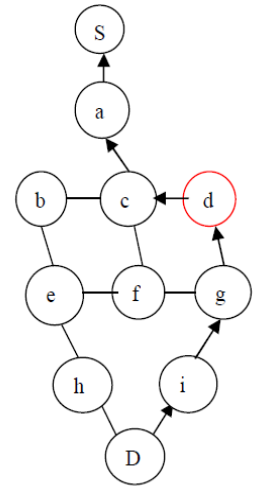

**Fig 8: RREP send by D in a reverse manner to S**

### 5.1.6 Step 6
Data delivery starts from S to D as DATA[S, a, c, d, g, i, D], is shown in figure 9. But d is malicious node and drops the packet in that route (Blackhole attack). So, we transfer the packet in a different safe route to reach the destination with the RREQ which reached second to D.

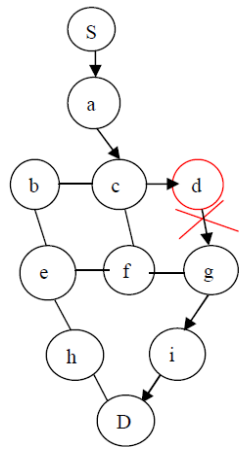

**Fig 9: Data delivery as DATA[S, a, c, d, g, i, D]**

## 6. CONCLUSION

SFTP algorithm shows better network performance by checking the effect of Blackhole attack on the network. By this we get a fault-free MANET which is adaptable to combat against the packet dropping attack. The three phases in SFTP algorithm provides robust security to the network against the malicious nodes.

In future, we will implement the SFTP algorithm. We also work on the performance parameters by simulating the results with a network simulator to reduce the packet delivery time, latency and increasing the throughput. We also add new security concepts by using the secure routing protocols in MANET.

## 7. REFERENCES

[1] Andel, T. R., and Yasinsac, A. 2007. Surveying Security Analysis Techniques in MANET Routing Protocols. IEEE Communications Surveys and Tutorials 9, 4 70-84.

[2] Abusalah, L., Khokhar, A., and Guizani, M. 2008. A Survey of Secure Mobile Ad Hoc Routing Protocols. IEEE Communications Surveys and Tutorials 10, 4 78-93.

[3] Bala, A., Bansal, M., and Singh, J. 2009. Performance Analysis of MANET under Blackhole Attack. First International Conference on Networks and Communications 141-145.

[4] Balakrishnan, V., Varadharajan, V., and Tupakula, U. K. 2006. Fellowship: Defense against Flooding and Packet Drop Attacks in MANET. 10$^{th}$ IEEE / IFIP Network Operations and Management Symposium (Apr. 2006).

[5] Bhalaji, N., and Shanmugam, A. 2009. Association Between Nodes to Combat Blackhole Attack in DSR based MANET. IFIP International Conference on Wireless and Optical Communications Networks (Apr. 2009).

[6] Burmester, M., and Medeiros, B. D. 2009. On the Security of Route Discovery in MANETs. IEEE Transaction on Mobile Computing 8, 9 (Sep. 2009), 1180-1188.

[7] Ford, R., and Howard, M. 2008. Security in Mobile Ad Hoc Networks. IEEE Security and Privacy 72- 75.

[8] Kannhavong, B., Nakayama, H., Nemoto, Y., and Kato, N. 2007. A Survey of Routing Attacks in Mobile Ad Hoc Networks. IEEE Wireless Communications (Oct. 2007) 85-91.

[9] Lima, M. N., Santos, A. L., and Pujolle, G. 2009. A Survey of Survivability in Mobile Ad Hoc Networks. IEEE Communications Surveys and Tutorials 11, 1 66-77.

[10] Lima, M. N., Silva, H. W., Santos, A. L., and Pujolle, G. 2008. Requirements for survivable routing in MANETs. 3$^{rd}$ International Symposium on Wireless Pervasive Computing (May. 2008), 441-445.

[11] Medadian, M., Yektaie, M. H. and Rahmani, A. M. 2009. Combat with Black Hole Attack in AODV routing protocol in MANET. First Asian Himalayas International Conference on Internet (Nov. 2009).

[12] Papadimitratos, P., and Haas, Z. J. 2006. Secure Data Communication in Mobile Ad Hoc Networks. IEEE Journalon Selected Area in Communications 24, 2 (Feb. 2006), 343-356.

[13] Purohit, N., Sinha, R., and Maurya, K. 2011. Simulation study of Black hole and Jellyfish attack on MANET using NS3. IEEE International Conference on Current Trends in Technology (Dec. 2011), 1-5.

[14] Sharma, N., and Sharma, A. 2012. The Black-hole node attack in MANET. Second International Conference on Advanced Computing and Communication Technologies 546-550.

[15] Tamilselvan, L., and Sankaranarayanan V. 2007. Prevention of Blackhole Attack in MANET. The 2$^{nd}$ International Conference on Wireless Broadband and Ultra Wideband Communications IEEE.

[16] Tsou, P., Chang, J., Lin, Y., Chao, H., and Chen, J. 2011. Developing a BDSR Scheme to Avoid Black Hole Attack Based on Proactive and Reactive Architecture in MANETs. ICACT (Feb. 2011), 755-760, ISBN= 978-89-5519-155-4.

[17] Zhang, X. Y., Sekiya, Y., and Wakahara, Y. 2009. Proposal of a Method to Detect Black Hole Attack in MANET. International Symposium on Autonomous Decentralized Systems (Mar. 2009).